\documentclass[mathleft
]{an}
\usepackage{times}
\usepackage[latin1]{inputenc}
\usepackage[dvips]{graphicx}
\usepackage[authoryear]{natbib}
\usepackage[english]{babel}
\usepackage{color}
\usepackage{longtable}
\usepackage{latexsym}       
\usepackage{url}            
\def\eg{\textit{e.\,g.}}                                      
\def\cf{\textit{cf.}}

\def\ros{{\it ROSAT}}
\def\ein{{\it Einstein}}
\def\chandra{{\it Chandra}}
\def\xmm{{\it XMM-Newton}}
\def\m31{\object{M~31}}

\newcommand{\ergcm}[1]{$\times 10^{#1}$ \hbox{erg cm$^{-2}$ s$^{-1}$}}
\newcommand{\ergs}[1]{$\times 10^{#1}$ \hbox{erg s$^{-1}$}}
\newcommand{\oergs}[1]{$10^{#1}$~erg~s$^{-1}$}
\newcommand{\oexpo}[1]{$10^{#1}$}
\newcommand{\mr}{\mathrm}
\def\XLPt{XMM\ LP-total}
\def\ga{\mathrel{\mathchoice {\vcenter{\offinterlineskip\halign{\hfil
$\displaystyle##$\hfil\cr>\cr\sim\cr}}}
{\vcenter{\offinterlineskip\halign{\hfil$\textstyle##$\hfil\cr
>\cr\sim\cr}}}
{\vcenter{\offinterlineskip\halign{\hfil$\scriptstyle##$\hfil\cr
>\cr\sim\cr}}}
{\vcenter{\offinterlineskip\halign{\hfil$\scriptscriptstyle##$\hfil\cr
>\cr\sim\cr}}}}}

\bibpunct{(}{)}{;}{a}{}{,} 

\begin{document}

\Pagespan{789}{}
\Yearpublication{2006}%
\Yearsubmission{2005}%
\Month{11}%
\Volume{999}%
\Issue{88}%

\title{Supersoft sources in M~31: Comparing the XMM-Newton Deep Survey, ROSAT and Chandra catalogues\,\thanks{Based on 
   observations obtained with XMM-Newton, an ESA science mission with 
   instruments and contributions directly funded by ESA Member States 
   and NASA.}}

\author{H. Stiele\inst{1}\fnmsep\thanks{Corresponding author:
  \email{hstiele@mpe.mpg.de}\newline}
 \and W.~Pietsch\inst{1} \and 
           F.~Haberl\inst{1} \and
	   V.~Burwitz\inst{1} \and
	  D.~Hatzidimitriou\inst{2,}\inst{3} \and 
J.~Greiner\inst{1}  
        }
\titlerunning{Supersoft sources in M~31}
\authorrunning{H. Stiele et al.}
\institute{Max-Planck-Institut f\"ur extraterrestrische Physik, Giessenbachstra{\ss}e,
           85748 Garching, Germany 
\and Department of Astrophysics, Astronomy and Mechanics, Faculty of
Physics, University of Athens, Panepistimiopolis, 15784 Zografos,
Athens, Greece
\and IESL, Foundation for Research and Technology, 71110 Heraklion, Greece
}

\received{??}
\accepted{??}
\publonline{later}


\abstract{%
To investigate the transient nature of supersoft sources (SSSs) in \m31, we compared SSS candidates of the \xmm\ Deep Survey, \ros\ PSPC~surveys and the \chandra\ catalogues in the same field.
We found 40 SSSs in the \xmm\ observations. While 12 of the \xmm\ sources were brighter than the limiting flux of the \ros\ PSPC survey, only two were detected with \ros\ $\sim$10~yr earlier. Five correlate with recent optical novae which explains why they were not detected by \ros.
The remaining 28 \xmm\ SSSs have fluxes below the \ros\ detection threshold. Nevertheless we found one correlation with a \ros\ source, which had significantly larger fluxes than during the \xmm\ observations. 
Ten of the \xmm\ SSSs were detected by \chandra\ with $<$1--$\sim$6~yr between the observations. Five were also classified as SSSs by \chandra.
Of the 30 \ros\ SSSs three were confirmed with \xmm, while for 11 sources other classifications are suggested. Of the remaining 16 sources one correlates with an optical nova. 
Of the 42 \chandra\ very-soft sources five are classified as \xmm\ SSSs, while for 22 we suggest other classifications. Of the remaining 15 sources, nine are classified as transient by \chandra, one of them correlates with an optical nova. 
These findings underlined the high variability of the sources of this class and the connection between SSSs and optical novae. Only three sources, were detected by all three missions as SSSs. Thus they are visible for more than a decade, despite their variability.}

\maketitle

\section{Introduction}
The class of supersoft sources (SSSs), was established by \ros\ 
and is based on observable characteristics. 
SSSs \linebreak show extremely soft spectra with equivalent blackbody temperatures of $\sim$15--80\,eV. The bolometric luminosities lie in the range of \oexpo{36}--\oergs{38}. The favoured model for these sources is that they are close binary systems with a white dwarf (WD) primary, burning hydrogen-rich matter at the surface. Symbiotic systems, which contain a white dwarf in a wide binary system, were also observed as SSSs \citep[][]{1997ARA&A..35...69K}.  
SSSs are often observed as transient X-ray sources \citep[][and references therein]{2000NewA....5..137G}. \citet[][hereafter PFF\linebreak 2005 and  PHS2007]{2005A&A...442..879P,2007A&A...465..375P} showed that many SSSs in \m31\ correlate with classical novae.

The observations of the `Deep \xmm\ Survey of \m31' \citep{2008xng..conf...23S} provide, together with archival observations, a full coverage of the D$_{25}$ ellipse of \m31\ with high spatial and spectral resolution. A description of the observations, the source detection procedure and the full source catalogue will be published in a separate paper (Stie\-le et al., in preparation). Here, we discuss the 40 SSSs contained in the Deep \xmm\ Survey catalogue (hereafter \XLPt). The SSSs were selected on the basis of their hardness ratios (HR1$<$0, HR2$-$EHR2 $<$$-0.96$ or HR2 not defined; energy bands: B1:0.2--0.5\,keV, B2:0.5--1.0\,keV, B3:1.0--2.0\,keV). The hardness ratios and errors are defined as:
\begin{equation}
\mr{HR}i = \frac{B_{i+1} - B_{i}}{B_{i+1} + B_{i}}\; \mbox{and}
\end{equation}
\begin{equation}
 \mr{EHR}i = 2  \frac{\sqrt{(B_{i+1} EB_{i})^2 + (B_{i} EB_{i+1})^2}}{(B_{i+1} + B_{i})^2},
\end{equation}
where $B_{i}$ and $EB_{i}$ denote count rates and corresponding errors in energy band {\it i}. By cross-correlating with our nova catalogue\footnote{\url{http://www.mpe.mpg.de/~m31novae/opt/m31/M31_table.html}} we found that 14 out of the 40 sources can be classified as X-ray counterparts of optical novae. The main properties of the four sources that are reported for the first time are given in Table~\ref{Tab:Nov}, while the remaining ten were reported in PFF2005, PHS2007 or \citet{2006IBVS.5737....1S}.

\begin{table}
\scriptsize
\begin{center}
\caption{Properties of four newly detected SSSs, which correlate with optical novae}
\begin{tabular}{lrrrrrr}
\hline\noalign{\smallskip}
\hline\noalign{\smallskip}
\multicolumn{1}{l}{Name} & \multicolumn{1}{l}{$t_{\mr{S}}^{*}$}& \multicolumn{1}{l}{$t_{\mr{E}}^{+}$}& \multicolumn{1}{l}{Dist$^{\&}$}& \multicolumn{1}{l}{Perr$^{\dagger}$}& \multicolumn{1}{l}{$T_{\mr{bb}}^{\ddagger}$}& \multicolumn{1}{c}{$L_{\mr{X}}^{\#}$}\\
\noalign{\smallskip}
\multicolumn{1}{l}{M31N} & \multicolumn{1}{l}{(d)}& \multicolumn{1}{l}{(d)}& \multicolumn{1}{c}{(\arcsec)}& \multicolumn{1}{c}{(\arcsec)}& \multicolumn{1}{l}{(eV)}& \multicolumn{1}{l}{(\oergs{37})}\\
\hline\noalign{\smallskip}
1997-10c & 982 & 1167 & 1.9 & 4.4 & 41 & 5.9\\
2005-01b & 535 & 1073 & 4.3 & 5.2 & 45 & 1.0\\
2005-01c & 703 &      & 0.9 & 1.8 & 40 & 12 \\
2005-09b & 299 &  690 & 0.6 & 3.5 & 35 & 54 \\
\noalign{\smallskip}
\hline
\noalign{\smallskip}
\end{tabular}
\label{Tab:Nov}
\end{center}
Notes:\\
$^{~*}~$: Time of first detection in X-rays (days after opt. outburst )\\
$^{~+}~$: Time of last detection in X-rays, (days after opt. outburst)\\
$^{~\&}~$: Distance between position of X-ray source and that of optical counterpart\\
$^{~\dagger}~$: 3$\sigma$ (99.7\%) positional error of the X-ray source\\
$^{~\ddagger}~$: Temperature derived from a blackbody fit to the X-ray spectrum\\
$^{~\#}~$: Unabsorbed 0.2--1.0\,keV luminosity, derived from the blackbody fit
\normalsize
\end{table}

To study the long term variability of the SSS population of \m31\ we performed cross-correlations with the \ros\ PSPC surveys \citep[][hereafter SHP97 and SHL2001]{1997A&A...317..328S,2001A&A...373...63S} and with the \chandra\ source catalogues \linebreak (\citealt{2002ApJ...577..738K}, \citealt{2002ApJ...578..114K}, \citealt{2006ApJ...643..356W}, \linebreak \citealt{2007A&A...468...49V}, \citealt{2004ApJ...610..247D}, hereafter \linebreak DKG2004). 

As the \ros\ PSPC observed \m31\ more than 10\,yr before \xmm\ EPIC, a comparison of the \ros\ to the \xmm\ detections probes the long-term variability of SSSs. Lists of SSSs detected in the \ros\ PSPC surveys are given by \citet{2000NewA....5..137G} and \citet{1999A&A...344..459K}. The selection of these \ros\ SSSs was based on similar selection criteria, as those used for the \xmm\ data, as the separation energies of the \ros\ bands were $\sim$0.5\,keV and $\sim$1.0\,keV.\@ We ignored all sources of the complimentary sample of \citet{1999A&A...344..459K} which were already classified as foreground stars or SNRs by SHP97. Thus the \ros\ sample contains 34 SSSs.  

A list of very soft sources detected with \chandra\ is given by DKG2004. The observations cover four isolated fields, located in the northern and southern disk and in the centre of \m31. DKG2004 developed an algorithm  to select SSSs based on the count rates found in the three energy bands: S, 0.1--1.1\,keV; M, 1.1--2\,keV; and H, 2--7\,keV. The most important difference to our \xmm\ study is the usage of only \emph{one} energy band below $\sim$1\,keV. That means that DKG2004 only use cuts that correspond to cuts in HR2 (and not HR1) for \xmm, to select SSSs. From \xmm\ HR2 cuts we know that -- no matter how low the HR2 threshold is chosen -- one cannot avoid to select foreground stars and SNRs \citep[][hereafter PFH2005]{2005A&A...434..483P}.

\section{Method}
\label{Sec:SSS_comp}
The analysis was carried out in two stages. First, we selected all SSSs from the \XLPt\ catalogue and cross-cor\-re\-la\-ted them with the full source catalogues obtained from \ros\ and \chandra\ observations described above, to investigate whether the \xmm\ sources had been detected in previous studies and their variability (Sec.\,\ref{SubSec:XMMROS}). Second, all sources classified as supersoft in the \ros\ PSPC surveys and as very soft in \chandra\ observations (SKG2004), respectively, were cross-correlated with the full \XLPt\ catalogue. These correlations not only allow us to study the variability of SSSs, but also to ascertain the selection power of the methods used in \ros\ and \chandra\ studies to separate different source classes (Secs.\ \ref{SubSec:ROSXMM} and \ref{SubSec:ChaXMM}). Obviously, sources that are assigned a supersoft state in both the \XLPt\ catalogue and in one or both of the \ros\ and \chandra\ catalogues, are recovered in both stages of the cross-correlation process. These sources are discussed in Sec.\,\ref{SubSec:XMMROS} and also mentioned in the Secs.\,\ref{SubSec:ROSXMM} and \ref{SubSec:ChaXMM} for completeness. 

\section{Results}
\label{Sec:Res}
\begin{figure}
\resizebox{\hsize}{!}{\includegraphics[clip]{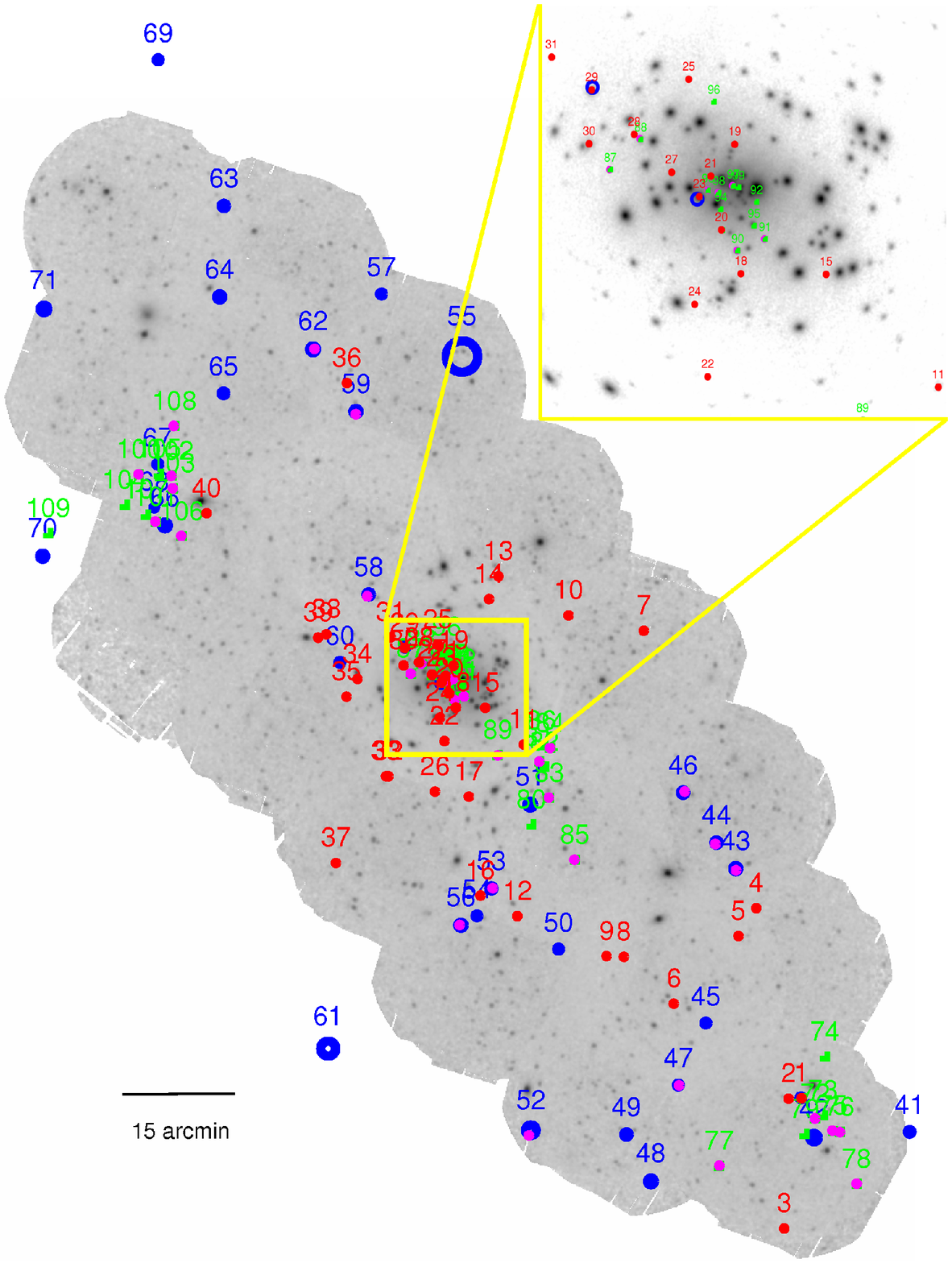}}
\caption{Image of the deep \xmm\ survey of \m31\ over plotted with the sources from Table \ref{Tab:SSS_overV}. SSSs from the \XLPt\ catalogue are marked in red, from DKG2004 in green and from the \ros\ PSPC surveys in blue. \chandra\ and \ros\ SSSs with \xmm\ counterpart not classified as SSS are marked in magenta. The image in the upper right corner shows a zoom-in to the central region of \m31, marked by the yellow box.}
\label{Figure1}
\end{figure}

Table \ref{Tab:SSS_overV} lists all examined sources (40 \xmm\ SSS candidates from the \XLPt\ catalogue, 34 \ros\ SSS candidates and 43 \chandra\ very soft sources from Table 1 of DKG2004) and is structured as follows: Column~2 indicates from which source list the examined source is\linebreak taken, while Col.\,3 provides the number (name) of the\linebreak  source in that list. The next three columns (4--6) give information about correlating sources from the other catalogues (and in some cases provide additional information obtained from studies with the same instrument). Sources that are observed as SSSs with more than one instrument are only listed once, reducing the \ros\ and \chandra\ lists to 31 and 38, respectively. In the last column (7) additional remarks are given. The positions of the sources are indicated in Fig.\,\ref{Figure1}.

\begin{table*}
\scriptsize
\begin{center}
\caption{Overview of SSS detected with \ros, \chandra\ and \xmm.}
\begin{tabular}{rcrrclclcl}
\hline\noalign{\smallskip}
\hline\noalign{\smallskip}
\multicolumn{1}{c}{Num} &\multicolumn{1}{c}{I$^{\&}$} &\multicolumn{1}{c}{Corr$^{\&}$} & \multicolumn{2}{c}{\xmm$^{+}$} & \multicolumn{2}{c}{\ros$^{\diamondsuit}$} & \multicolumn{2}{c}{\chandra$^{\#}$} & Remarks$^{\ddagger}$ \\
\multicolumn{1}{c}{(1)} &\multicolumn{1}{c}{(2)} &\multicolumn{1}{c}{(3)} &\multicolumn{2}{c}{(4)} &\multicolumn{2}{c}{(5)} &\multicolumn{2}{c}{(6)}&\multicolumn{1}{c}{(7)}  \\
\noalign{\smallskip}\hline\noalign{\smallskip}
1         & X & 69      &          &$<$SSS$>$     & SI 18, SII 27                          &$<$SSS$>$     &  s2-26 (D,v)                        & $<$SSS$>$     &      \\
2         & X & 92      &          &$<$SSS$>$     &                                        &              &                                     &               &   M31N~2005-09b\\
3         & X & 97      &          &$<$SSS$>$     &                                        &              &                                     &               &             \\
4$^{*}$   & X & 147     &          &$<$SSS$>$     &                                        &              &                                     &               &   no Chandra   \\
5         & X & 183     &          &$<$SSS$>$     &                                        &              &                                     &               &            \\
6         & X & 342     &          &$<$SSS$>$     &                                        &              &                                     &               &       \\
7         & X & 408     &          &$<$SSS$>$     &                                        &              &                                     &               &   no Chandra    \\
8         & X & 454     &          &$<$SSS$>$     &                                        &              &                                     &               &       \\
9         & X & 511     &          &$<$SSS$>$     &                                        &              &                                     &               &       \\
10        & X & 617     &          &$<$SSS$>$     &                                        &              &                                     &               &   no Chandra  \\
11$^{*}$  & X & 748     &          &$<$SSS$>$     &                                        &              &                                     &               &   M31N~2001-10f\\
12        & X & 764     &          &$<$SSS$>$     &                                        &              &                                     &               &   M31N~2005-01b\\
13        & X & 821     &          &$<$SSS$>$     &                                        &              &                                     &               &             \\
14        & X & 857     &          &$<$SSS$>$     & SII 156$^{\dagger}$                    &$<$fg Star$>$ &                                     &               &       \\
15        & X & 871     &          &$<$SSS$>$     & SI 160$^{\dagger}$                     &              &                                     &               &   M31N~1997-10c\\
16        & X & 887     &          &$<$SSS$>$     &                                        &              &                                     &               &     \\
17$^{*}$  & X & 934     &          &$<$SSS$>$     &                                        &              &                                     &               &   M31N~2007-06b\\
18        & X & 993     &          &$<$SSS$>$     &                                        &              & Ka 55                               &               &   M31N~1994-09a \\
19$^{*}$  & X & 1\,006  &          &$<$SSS$>$     &                                        &              & Ka 67, r2-60 (D,t, W6)              & $<$SSS$>$     &   M31N~2000-07a\\
20        & X & 1\,025  &          &$<$SSS$>$     &                                        &              & Ka 86, r2-65 (D,t, W6)              & $<$SSS$>$     &    \\
21        & X & 1\,046  &          &$<$SSS$>$     &                                        &              &                                     &               &   M31N~1999-10a\\
22        & X & 1\,051  &          &$<$SSS$>$     &                                        &              &                                     &               &   M31N~1997-08b\\
23$^{*}$  & X & 1\,061  &          &$<$SSS$>$     & SI 208, SII 203                        &$<$SSS$>$     & Ka 100, r2-12 (K,r,v, W, D,v), VG 46& $<$SSS$>$     &   217s period$^{\clubsuit}$, TF 69\\
24        & X & 1\,069  &          &$<$SSS$>$     &                                        &              &                                     &               &     \\
25        & X & 1\,076  &          &$<$SSS$>$     &                                        &              & Ka 106                              &               &   M31N~1996-08b  \\
26        & X & 1\,087  &          &$<$SSS$>$     &                                        &              &                                     &               &      \\
27        & X & 1\,100  &          &$<$SSS$>$     &                                        &              & Ka 111, r2-63 (W, D,t, W6),         & $<$SSS$>$     &   M31N~1995-11c \\
28        & X & 1\,144  &          &$<$SSS$>$     &                                        &              & r3-128 (W6)                         &               &    \\
29$^{*}$  & X & 1\,194  &          &$<$SSS$>$     & SI 235, SII 235                        &$<$SSS$>$     & Ka 135, r3-8 (K,v, W, W6), VG 214   & $<$SSS$>$     &   $^{\infty}$, TF 87 \\
30$^{*}$  & X & 1\,195  &          &$<$SSS$>$     &                                        &              & r3-126 (W, W6)                      &               &   865.5s period$^{\spadesuit}$\\
31        & X & 1\,236  &          &$<$SSS$>$     &                                        &              &                                     &               &   M31N~1998-06a \\
32        & X & 1\,242  &          &$<$SSS$>$     &                                        &              &                                     &               &     \\
33$^{*}$  & X & 1\,250  &          &$<$SSS$>$     &                                        &              &                                     &               &   time variable \\
34        & X & 1\,325  &          &$<$SSS$>$     &                                        &              &                                     &               &          \\
35        & X & 1\,355  &          &$<$SSS$>$     &SI 262$^{\dagger}$, SII 269$^{\dagger}$ &$<$SNR$>$     &                                     &               &       \\
36        & X & 1\,356  &          &$<$SSS$>$     &                                        &              &                                     &               &   no Chandra\\
37$^{*}$  & X & 1\,381  &          &$<$SSS$>$     &                                        &              &                                     &               &   no Chandra\\
38$^{*}$  & X & 1\,416  &          &$<$SSS$>$     &                                        &              &                                     &               &   \\
39        & X & 1\,435  &          &$<$SSS$>$     &                                        &              &                                     &               &    \\
40$^{*}$  & X & 1\,675  &          &$<$SSS$>$     &                                        &              &                                     &               &   M31N~2005-01c\\
41        & R & SI 3    &          &              &                                        &$<$SSS$>$     &                                     &               &   outside FoV\\
42        & R & SI 12   &          &              &                                        &$<$SSS$>$     &                                     &               &   \\
43        & R & SI 35   & 188      &$<$fg Star$>$ & SII 43                                 &$<$SSS$>$     &                                     &               &      \\
44        & R & SI 39   & 240      &$<$hard$>$    &                                        &$<$SSS$>$     &                                     &               &             \\
45        & R & SI 45   &          &              & SII 51                                 &$<$SSS$>$     &                                     &               &   \\
46        & R & SI 58   & 304      &$<$GCl$>$     & SII 60                                 &$<$SSS$>$     &                                     &               &         \\
47        & R & SI 62   & 325      &$<$fg Star$>$ &                                        &$<$SSS$>$     &                                     &               &    \\
48        & R & SI 78   &          &              &                                        &$<$SSS$>$     &                                     &               &   \\
49        & R & SI 88   &          &              &                                        &$<$SSS$>$     &                                     &               &   \\
50        & R & SI 114  &          &              &                                        &$<$SSS$>$     &                                     &               &  \\
51        & R & SI 128  &          &              &                                        &$<$SSS$>$     &                                     &               &  \\
52        & R & SI 129  & 737      &$<$fg Star$>$ & SII 123                                &$<$SSS$>$     &                                     &               &       \\
53        & R & SI 156  & 842      &$<$fg Star$>$ & SII 151                                &$<$SSS$>$     &                                     &               &    \\
54        & R & SI 171  &          &              &                                        &$<$SSS$>$     &                                     &               &   \\
55        & R & SI 183  & \multicolumn{2}{c}{10 XMM ctps$^{\natural}$} & SII 185           &$<$SSS$>$     &                                     &               &  \\
56        & R & SI 185  & 969      &$<$SNR$>$     &  SII 186                               &$<$SSS$>$     & s1-84 (W)                           &               &      \\
57        & R & SI 245  &          &              &                                        &$<$SSS$>$     &                                     &               &    \\
58        & R & SI 252  & 1\,297   &$<$hard$>$    & SII 258                                &$<$SSS$>$     &                                     &               &       \\
59        & R & SI 259  & 1\,331   &$<$fg Star$>$ &                                        &$<$SSS$>$     &                                     &               &       \\
60        & R & SI 268  &          &              &                                        &$<$SSS$>$     &                                     &               &   M31N 1990-09a$^{\heartsuit}$  \\
61        & R & SI 271  &          &              & SII 282                                &$<$SSS$>$     &                                     &               &   outside FoV\\
62        & R & SI 280  & 1\,442   &fg Star       & SII 287                                &$<$SSS$>$     &                                     &               &        \\
63        & R & SI 307  &          &              &                                        &$<$SSS$>$     &                                     &               &    \\
64        & R & SI 309  &          &              & SII 324                                &$<$SSS$>$     &                                     &               &  \\
65        & R & SII 322 &          &              &                                        &$<$SSS$>$     &                                     &               &   \\
66        & R & SI 330  &          &              &                                        &$<$SSS$>$     &                                     &               &    \\
67        & R & SI 335  &          &              &                                        &$<$SSS$>$     &                                     &               &    \\
68        & R & SI 341  &          &              &                                        &$<$SSS$>$     &                                     &               &    \\
69        & R & SI 342  &          &              &                                        &$<$SSS$>$     &                                     &               &    outside FoV\\
70        & R & SI 374  &          &              &                                        &$<$SSS$>$     &                                     &               &    outside FoV\\
71        & R & SI 376  &          &              &                                        &$<$SSS$>$     &                                     &               &    \\
72        & C & s2-7    & 52       &$<$fg Star$>$ &                                        &              & (D)                                 & $<$SSS$>$     &   \\
73        & C & s2-10   &          &              &                                        &              & (D)                                 & $<$QSS$>$     &   \\
74        & C & s2-27   &          &              &                                        &              & (D)                                 & $<$QSS$>$     &    \\
75        & C & s2-28   & 32       &              &                                        &              & (D)                                 & $<$QSS$>$     &               \\
76        & C & s2-29   & 23       &$<$fg Star$>$ &                                        &              & (D)                                 & $<$fg Star$>$ & $^{\sharp}$  \\
77        & C & s2-37   & 237      &$<$fg Star$>$ & SI 40, SII 47                          &              & (D)                                 & $<$fg Star$>$ & $^{\flat}$   \\
78        & C & s2-46   & 13       &$<$fg Star$>$ &                                        &              & (D,v)                               & $<$fg Star$>$ & $^{\sharp}$  \\
\noalign{\smallskip}
\hline
\noalign{\smallskip}
\end{tabular}
 \label{Tab:SSS_overV}
\end{center}
\normalsize
\end{table*}

\begin{table*}
\addtocounter{table}{-1}
\scriptsize
\begin{center}
\caption{Overview of SSS detected with \ros, \chandra\ and \xmm. (continued)}
\begin{tabular}{rcrrclclcl}
\hline\noalign{\smallskip}
\hline\noalign{\smallskip}
\multicolumn{1}{c}{Num} &\multicolumn{1}{c}{I$^{\&}$} &\multicolumn{1}{c}{Corr$^{\&}$} & \multicolumn{2}{c}{\xmm$^{+}$} & \multicolumn{2}{c}{\ros$^{\diamondsuit}$} & \multicolumn{2}{c}{\chandra$^{\#}$} & Remarks$^{\ddagger}$ \\
\multicolumn{1}{c}{(1)} &\multicolumn{1}{c}{(2)} &\multicolumn{1}{c}{(3)} &\multicolumn{2}{c}{(4)} &\multicolumn{2}{c}{(5)} &\multicolumn{2}{c}{(6)} &\multicolumn{1}{c}{(7)} \\
\noalign{\smallskip}\hline\noalign{\smallskip}                                           
79  & C & s2-62   &          &               &                                    &                   & (D,t)                                             & $<$QSS$>$     &    \\
80  & C & s1-18   &          &               &                                    &                   & (D,v,t)                                           & $<$SSS$>$     &    \\
81  & C & s1-20   & 696      & fg Star       &  SI 121, SII 112                   &  $<$fg Star$>$    & (D)                                               & $<$fg Star$>$ & $^{\flat}$         \\
82  & C & s1-27   &          &               &                                    &                   & (D,v,t)                                           & $<$QSS$>$     &    \\
83  & C & s1-41   & 673      & $<$Gal$>$     &                                    &                   & (D)                                               & $<$GlC$>$     & $^{\flat}$      \\
84  & C & s1-42   & 668      & SNR           & SI 116                             &  SNR              & (D)                                               & SNR           &  $^{\sharp}$            \\
85  & C & s1-45   & 603      & $<$fg Star$>$ & SI 107, SII 99                     &  $<$fg Star$>$    &  (D,W)                                            & $<$fg Star$>$ & $^{\flat}$  \\
86  & C & s1-69   &          &               &                                    &                   & (D,t)                                             & $<$SSS$>$     &    \\
87  & C & r3-11   & 1\,172   & $<$hard$>$    &                                    &                   & (K, D), VG 161                                    & $<$QSS$>$     &    \\
88  & C & r3-115  & 1\,136   & $<$XRB$>$     &                                    &                   & (W, D,t, W6), Ka 125, VG 128 (t)                  & $<$SSS$>$     &   \\
89  & C & r3-122  & 826      & $<$fg Star$>$ &  SI 147                            &  $<$fg Star$>$    & (D)                                               & $<$fg Star$>$ & $^{\sharp}$   \\
90  & C & r2-19   & 1\,000   &               &                                    &                   & (K,f, W, D), Ka 63, VG 72 (t)                     & $<$QSS$>$     &        \\
91  & C & r2-42   & 960      & $<$fg Star$>$ &  SI 181, SII 182                   &                   & (K,f, W, D), Ka 43, VG 69                         & $<$QSS$>$     &   \\
92  & C & r2-54   &          &               &                                    &                   & (D)                                               & $<$SSS$>$     &   \\
93  & C & r2-56   & 1\,050   & SNR           &                                    &                   & (K,p, D), VG 36,                                  & SNR           &$^{\sharp}$        \\
94  & C & r2-61   &          &               &                                    &                   & (D,t)                                             & $<$SSS$>$     &   M31N2000-08a$^{\heartsuit}$\\
95  & C & r2-62   &          &               &                                    &                   & (D,t)                                             & $<$QSS$>$     &   \\
96  & C & r2-66   &          &               &                                    &                   & (D,t)                                             & $<$QSS$>$     &   \\
97  & C & r1-9    & 1\,010$^{\dagger}$   & $<$XRB$>$     &                                    &                   & (K,r,v,t, W, D,v,t, W6)                           & $<$QSS$>$     &   unresolved  \\
98  & C & r1-25   & 1\,034   & $<$XRB$>$     &                                    &                   & (K, W, D), Ka 89, VG 23 (t)                       & $<$SSS$>$     &     \\
99  & C & r1-35   &          &               &                                    &                   & (D,t)                                             & $<$SSS$>$     &   M31N1995-09b$^{\heartsuit}$ \\
100 & C & n1-2    & 1\,806   & $<$fg Star$>$ &                                    &                   & (D)                                               & $<$SSS$>$     &   \\
101 & C & n1-8    & 1\,773   & $<$fg Star$>$ &                                    &                   & (D)                                               & $<$QSS$>$     &   \\
102 & C & n1-13   & 1\,747   & $<$fg Star$>$ &                                    &                   & (D)                                               & $<$QSS$>$     &   \\
103 & C & n1-15   & 1\,742   & $<$fg Star$>$ &  SI 327                            &                   & (D)                                               & $<$QSS$>$     &   \\
104 & C & n1-26   &          &               &                                    &                   & (D,v,t)                                           & $<$QSS$>$     &   \\
105 & C & n1-29   &          &               &                                    &                   & (D)                                               & $<$QSS$>$     &   \\
106 & C & n1-31   & 1\,721   & $<$hard$>$    &                                    &                   & (D)                                               & $<$QSS$>$     &       \\
107 & C & n1-46   &          &               &                                    &                   & (D)                                               & $<$QSS$>$     &   \\
108 & C & n1-48   & 1\,741   & $<$SNR$>$     &                                    &                   & (D)                                               & $<$SSS$>$     &      \\
109 & C & n1-66   &          &               & {\color{white},SI 208, SII 203}    &                   & (D){\color{white},Ka 100, r2-12 (K,r,v, W, D,v), VG } &  $<$QSS$>$&   outside FoV\\
\noalign{\smallskip}                                           
\hline                                                         
\noalign{\smallskip}
\end{tabular}
\end{center}
Notes:
$^{ {\&}~}$: Correlated SSS taken from the \XLPt\ catalogue, \ros\ SSS lists \citet{2000NewA....5..137G}, \citet{1999A&A...344..459K} (given is the source name from SHP97 or SHL2001) and very soft sources from Table~1 of DKG2004; The used lists are marked in the instrument column (I): X: \xmm, C: \chandra, R: \ros \\ 
$^{ {+}~}$: Sources and classification from the \XLPt\ catalogue (\citeauthor{Stiele} et al., in preparation)\\
$^{ {\#}~}$: Sources from \chandra\ catalogues: D: \citet{2004ApJ...610..247D}, K: \citet{2002ApJ...577..738K}, VG: \citet{2007A&A...468...49V}, W: \citet{2004ApJ...609..735W}, W6: \citet{2006ApJ...643..356W}, Ka: \citet{2002ApJ...578..114K}, v: variable, t: transient, r: \ros\ HRI source, f: foreground star, p: planetary nebula; and classification of \chandra\ sources from DKG2004; $^{ {\infty}~}$: apart from r3-8 which was classified as SSS by \citet{2002ApJ...577..738K}; $^{ {\sharp}~}$: listed as $<$SSS$>$ or $^{ {\flat}~}$: $<$QSS$>$ in Table~1 of DKG2004\\ 
$^{ {\diamondsuit}~}$: Sources from \ros\ catalogues SI: SHP97, SII: SHL2001 and classification of \ros\ sources from \citet{2000NewA....5..137G}, \citet{1999A&A...344..459K}, SHP97 and SHL2001\\
$^{ {\ddagger}~}$: ``outside FoV" means \ros\ or \chandra\ source is located outside the area covered by the Deep \xmm\ Survey of \m31; ``no Chandra" means that the location of the source was not covered by any \chandra\ observation (included in the used literature), TF: \citet{1991ApJ...382...82T}\\
$^{ {*}~}$: \xmm\ SSSs with 0.2--4.5\,keV flux above \ros\ PSPC detection threshold\\
$^{ {\dagger}~}$: chance coincidence; $^{ {\clubsuit}~}$:\citet{2008ApJ...676.1218T}; $^{ {\spadesuit}~}$:\citet{2001A&A...378..800O}; $^{ {\heartsuit}~}$: more details in PFF2005 or PHS2007; $^{ {\natural}~}$: counterparts
\normalsize
\end{table*}

\subsection{Correlating \textit{XMM-Newton} SSSs to the \textit{ROSAT} PSPC surveys and to \textit{Chandra} catalogues}
\label{SubSec:XMMROS}
Three \xmm\ SSSs correlate with sources that are also classified as supersoft in the \ros\ and \chandra\ catalogues. Two of them are located in the central field of \m31. The first one (\#29) was classified as an SSS by \citet{2000NewA....5..137G}, but not by SHP97 or SHL2001. The second one (\#23) has been known since \ein. Hence it has been visible for over 25 years. \citet{2008ApJ...676.1218T} found that this source varies with a period of 217\,s. The third source (\#1) is located in the southernmost field of the \xmm\ survey and correlates with [SHL2001]27 and s2-26 (DKG2004). [SHL2001]27 was detected in the \ros\ PSPC surveys with $F_{\mr{x,SHP97}}\!\approx\!5.19\!\times\!10^{-14}$\,erg cm$^{-2}$ s$^{-1}$ and $F_{\mr{x,SHL2001}}\!\approx\!4.56$\ergcm{-14}, which is a factor $\sim$22\,--\,25 higher than the fluxes derived from the\linebreak \xmm\ observations. DKG2004 detected s2-26 in\linebreak only one of three \chandra\ observations obtained between 2000 and 2001 and hence classified the source as variable. With \xmm\ the source was detected in all three observations\footnote {Observations from 2006 June 28, 2007 July 24 and 2008 January 2.} covering the position and did not show significant variability. Detection of this source in our survey suggests that it has been active for about 20 years.

Another three \xmm\ SSS candidates (\#19, \#20, \#27) have counterparts in the \chandra\ catalogues, that were classified as \chandra\ SSSs \citep[DKG2004,][]{2002ApJ...577..738K}.

Of the 40 SSS candidates found in the \xmm\ observations 12 have 0.2--4.5\,keV fluxes above the \ros\ PSPC detection threshold of $\sim\!5.3$\ergcm{-15}. They are marked with an ``$*$" in Table \ref{Tab:SSS_overV}. Nevertheless 10 of them do not have \ros\ counterparts. One of them (\#30) is a supersoft transient that shows variability with a periodicity of 865.5\,s \citep[][]{2001A&A...378..800O}. Another one (\#33) was detected in only one \xmm\ observation. Upper-limits to its flux obtained in three other \xmm\ observations indicate variability by a factor of $F_{\mr{var}}\!\ga\!6.9$ ($\sigma\!=\!2.8$).\@ Another six sources correlate with recent optical novae, which explains why they were not detected by \ros.\@  
The \ros\ non-detection of the remaining two sources implies a much weaker source flux at the time of the \ros\ observations. Therefore both sources are likely to be transient or at least variable.

Among the remaining 28 \xmm\ SSSs that have fluxes below the \ros\ detection threshold, we detected four correlations of which three, however, have to be considered as chance coincidences: In two cases a second (not supersoft) \xmm\ source correlates with the \ros\ counterparts. The third \xmm\ SSS correlates with an optical nova, that was detected in the optical $\sim$6\,yr after the \ros\ observations.

Of the 30 \xmm\ SSSs that do not have \chandra\ counterparts, five lie in regions that were not covered in the \chandra\ catalogues.\@ Another ten sources correlate with optical novae, which either had their optical outburst after the \chandra\ observations were obtained (so they could not be visible during the \chandra\ observations), or for which \chandra\ observations, if at all, can only provide upper limits (PFF2005 and PHS2007). In addition, source \#33 was found to be variable in our \xmm\ survey.

\subsection{Correlating \textit{ROSAT} SSSs to the \textit{XMM-Newton} Deep Survey catalogue} 
\label{SubSec:ROSXMM}
Of the 34 \ros\ SSS candidates four are located outside the field of \m31\ covered with \xmm\ observations. 
11 correlate with \XLPt\ sources with other classifications (see Table \ref{Tab:SSS_overV}). 
Source [SHP97]~183 correlated with ten \xmm\ sources, due to the large position error, rendering identification with a single source impossible. 

We examined the variability of the correlated sources between the \ros\ and \xmm\ observations following the method used in \citet{2008A&A...480..599S}. 
Many of the sources show rather low variability and are classified as\linebreak foreground star, SNR, or background galaxy candidates\linebreak from \xmm\ data. As sources of the afore mentioned classes (excluding flaring foreground stars) are expected to be hardly variable on time scales of several years (in contrast to SSSs), the detected lack of variability is consistent with the \xmm\ classifications.  

Sixteen \ros\ SSSs are left without a corresponding source from the \XLPt\ survey. One ([SHP97]~268) source correlates with the optical nova M31N 1990-09a \linebreak (PFF2005). The remaining 15 sources have to be classified as transient or at least highly variable and may well represent the SSS phase of optical novae that have not been detected in the optical. In the years before the \ros\ observations, there were no systematic searches or monitoring campaigns for optical novae in \m31. 
Hence, the number of known optical novae per year was very low (\citeauthor{Pietsch}, this issue).
          
\subsection{Correlating \textit{Chandra} very-soft sources to the \textit{XMM-Newton} Deep Survey catalogue} 
\label{SubSec:ChaXMM}
Table~1 of DKG2004 lists 43 very soft sources detected with \chandra, of which 20 are classified as SSSs. For five of them Table~3 of DKG2004 already lists correlations with known SNRs or foreground stars. This information is taken into account for the entries in the \chandra\ classification column of Table \ref{Tab:SSS_overV}.\@ The \xmm\ counterparts of 10 \chandra\ SSSs, do not show a supersoft spectrum (in the \xmm\ observations) and thus received a different \xmm\ source classification (see Table \ref{Tab:SSS_overV}). Two sources are especially interesting as they are classified as X-ray binary (XRB) candidates in the \XLPt\ catalogue. The first one (r3-115) has been noted to have an uncommon be\-hav\-iour and was discussed in detail in PFF2005. In early X-ray observations\footnote{\chandra: 1999 Oct 13, 1999 Dec 11, 1999 Dec 27, 2000 Jan 29, 2000 Feb 16, 2000 Jul 29, 2000 Aug 27, 2001 Jun 10 and \xmm: 2000 Jun 25, 2000 Dec 27, 2001 Jun 29} that transient source shows a supersoft spectrum. As the source luminosity increases, its spectrum becomes hard (in the \xmm\ observation from 2002 January 06). Based on its hard spectrum and transient behaviour the source was classified as an XRB candidate. The soft to hard transition suggests a black hole (BH) primary (PFF2005), although an optical nova or a symbiotic star cannot be excluded. The second \chandra\ SSS (r1-25) with an \xmm\ counterpart classified as XRB candidate, shows a very similar behaviour. It was detected with an X-ray 0.3\,--\,7\,keV luminosity of $\sim\!5$\ergs{35}.\@ HR1$_{\mr{kong}}\!=\!-0.79$, HR2$_{\mr{kong}}\!=\!-1.00$ indicate a supersoft spectrum \citep{2002ApJ...577..738K}. 
The source was not visible in the 2000\,--\,2002 \xmm\ observations of the centre of \m31.\@ A correlating source was detected in observations of July\linebreak 2004, with an 0.2--4.5\,keV luminosity of $\sim3.7\!\times$10$^{36}$\,erg s$^{-1}$ and a hard spectrum. The \xmm\ long-term variability analysis  \citep{2008A&A...480..599S} gives a maximum variability factor of $F_{\mr{var}}\!=\!9.35$ and $\sigma_{\mr{var}}\!=\!28.02$. \citet{2007A&A...468...49V} reported a \chandra\ detection of a correlating source ([VG2007]~23) in observation 4\,682, taken on 2004 May 23, which is about two months before the 2004 \xmm\ observations, at a luminosity of 1.26\ergs{37} and found a variability factor of 50.3.\@ An optical source located within 1\,\farcs2 of the \xmm\ source position is listed as a `\textsl{regular or semi-regular red variable}' in \citet{FRSB06}. 

From the 23 quasi-soft sources (QSSs) of Table~1 in DKG2004 (4 with correlations in Table~3 of DKG2004), 13 correlate with \xmm\ sources. The correlation with r1-9 has to be regarded as chance coincidence as \xmm\ cannot resolve this source, which is located in direct vicinity of the central source of \m31\ and a nearby XRB. In Table~3 of DKG2004 the correlation of s1-41 with the globular cluster candidate B251 (RBC V3.5) is indicated. \citet{2009AJ....137...94C} found that B251 is more likely to be a background galaxy than a globular cluster candidate.  
 
\section{Discussion and conclusions}
Of the 40 SSSs detected with \xmm\ only three \linebreak sources are visible for at least one decade. The additional six sources that were visible in \xmm\ and \chandra\ observations were all located in the central area of \m31. The \chandra\ and \xmm\ observations of that area were taken at about the same time (within several weeks to a few months, whereas in the outer regions there is at least a five year gap between the \chandra\ and \xmm\ observations). From all \xmm\ SSSs 12 have a flux above the \ros\ detection threshold. Nevertheless only two were detected in the \ros\ PSPC surveys. These findings underline the \emph{long} term variability of the class of SSSs \citep[\cf\ ][]{2004ApJ...610..261G}. 

From the 34 \ros\ SSSs (selected from \citealp{2000NewA....5..137G} and \citealp{1999A&A...344..459K}) four are outside the field observed with \xmm. We assigned other source types to the \xmm\ counterparts of 11 \ros\ sources, which means that 19 \ros\ SSS candidates are left. Subtracting the three SSSs that were confirmed by \xmm\ and the correlation with an optical nova, there are 15 sources left which must be considered as transient or at least variable. Due to the few optical novae, observed in the years before the \ros\ observations, several of these 15 sources may be the X-ray counterpart of a nova, where the optical outburst has been missed.\@  

DKG2004 report on 20 \chandra\ SSSs, of which three are classified as foreground stars and two as SNRs. From the remaining sources two \xmm\ counterparts are classified as foreground star candidates, one as SNR candidate and two as XRB candidates. The latter two sources are very interesting as they were found as SSSs in \chandra\ observations, but showed a ``hard" spectrum in \xmm\ observations. This indicates a transition from a supersoft to a hard spectral state, which is consistent with the behaviour of BH XRBs. However other source types, like \eg\ symbiotic stars, cannot be excluded. Five \chandra\ sources do not have counterparts in the \xmm\ observations. The fact that half of the \chandra\ SSSs are not \xmm\ SSSs underlines the missing selection sensitivity in the \chandra\ studies, as only one band below $\sim$1\,keV was used. Of the 23 \chandra\ quasi-soft sources (DKG2004), about half (12) have counterparts in the \xmm\ Deep Survey observations. However none of these 12 \xmm\ sources had hardness ratios consistent with SSSs.

In conclusion our comparative study of SSS candidates in \m31\ detected with \ros, \chandra\ and \xmm\ demonstrated that strict selection criteria have to be applied to securely select SSSs. It also underlined the high variability of the sources of this class and the connection between SSSs and optical novae.

\acknowledgements
We thank the referee for his very constructive report, which helped to substantially improve the paper.
The XMM-Newton project is supported by the
Bundesministerium f\"ur Wirtschaft und Technologie/Deutsches Zentrum
f\"ur Luft- und\linebreak Raumfahrt (BMWI/DLR,\,FKZ\,50\,OX\,0001) and the Max-Planck
Society. HS acknowledges support by the
Bundesministerium f\"ur Wirt\-schaft und Technologie/Deutsches Zentrum
f\"ur Luft- und\linebreak Raumfahrt (BMWI/DLR,\,FKZ\,50\,OR\,0405).

\bibliographystyle{aa}
\bibliography{/home/hstiele/MyThesis/papers,/home/hstiele/data1/papers/my1990,/home/hstiele/data1/papers/my2000,/home/hstiele/data1/papers/my2001,/home/hstiele/data1/papers/catalog,/home/hstiele/data1/papers/my2007,/home/hstiele/data1/papers/my2008}

\end{document}